\documentclass[aps,prb,twocolumn,superscriptaddress]{revtex4}

\usepackage{amsmath,amssymb}
\usepackage{graphicx}
\usepackage{epstopdf}
\usepackage{bm}
\usepackage{color}
\usepackage{mhchem}

\newcommand{\slr}{$T_1^{-1}$}
\newcommand{\slrt}{$(T_1T)^{-1}$}

\newcommand{\lsco}{La$_{2-x}$Sr$_x$CuO$_4$}
\newcommand{\lesco}{La$_{1.8-x}$Eu$_{0.2}$Sr$_x$CuO$_4$}
\newcommand{\lscos}{La$_{1.93}$Sr$_{0.07}$CuO$_4$}

\begin{document}


\title{$^{139}$La NMR investigation in underdoped \lscos}


\author{S.-H. Baek}
\email[]{sbaek.fu@gmail.com}
\affiliation{IFW-Dresden, Institute for Solid State Research, PF
270116, 01171 Dresden, Germany}
\author{A. Erb}
\affiliation{Walther-Mei{\ss}ner-Institut, Bayerische Akademie der Wissenschaften,
Walther-Mei{\ss}ner-Stra{\ss}e 8, D-85748 Garching, Germany}
\author{B. B\"{u}chner}
\affiliation{IFW-Dresden, Institute for Solid State Research,
PF 270116, 01171 Dresden, Germany}
\author{H.-J. Grafe}
\affiliation{IFW-Dresden, Institute for Solid State Research, PF
270116, 01171 Dresden, Germany}
\date{\today}

\begin{abstract}
We report $^{139}$La and $^{63}$Cu nuclear magnetic and quadrupole
resonance (NMR/NQR) studies in an underdoped \lscos\ single
crystal, focusing on the $^{139}$La NMR in the normal state. We
demonstrate that the local structural distortions in the
low-temperature orthorhombic structure cause the tilting of the
direction of the electric field gradient (EFG) at the nuclei from
the $c$ axis, resulting in two NMR central transition spectra at
both the $^{139}$La and $^{63}$Cu nuclei in an external field.
Taking into account the tilt angle of the EFG, the temperature
dependence of the $^{139}$La spectra allowed us to determine the
$^{139}$La Knight shift and the structural order parameter.  The
angle and temperature dependence of the $^{139}$La spectrum is in
perfect agreement with the macroscopic average structure and
proves a displacive transition. The $^{139}$La nuclear
spin-lattice relaxation rates, \slr, suggest that \lscos\
undergoes a gradual change to a temperature-independent paramagnetic
regime in the high temperature region. Both the spectra and \slr\
of the $^{139}$La as a function of temperature reveal a sharp
anomaly around $T_S=387(1)$ K, implying a first-order-like
structural transition, and a dramatic change below $\sim 70$ K
arising from collective glassy spin freezing.
\end{abstract}




\maketitle

\section{Introduction}

\lsco\ (LSCO) is a single-layered cuprate where the CuO$_2$ planes
are separated by a La$_{2-x}$Sr$_x$O$_2$ block layer. A mismatch
between the CuO$_2$ planes and the interlayer causes a structural
transition from a high temperature tetragonal (HTT) to a
low-temperature orthorhombic (LTO) phase,\cite{keimer92} where the
CuO$_6$ octahedra are tilted from the $c$ axis resulting in a
buckling pattern of the CuO$_2$ plane.
Further partial substitution of a rare earth element (Eu or Nd) at
the La site, or replacement of Sr by Ba induces another structural
transition in \lsco\ from the LTO to a low-temperature tetragonal
(LTT) phase. In the LTT structure, the local tilt of the CuO$_6$
octahedra persists but its direction is rotated by $45^\circ$ and
alternates by $90^\circ$ in neighboring planes yielding the
macroscopic tetragonal symmetry. Such a local structural
distortion as a function of doping has a large influence on the
magnetic and superconducting
phases.\cite{axe94,buchner94,bianconi96,suh99,klauss00} In
particular, the LTT phase seems to be a prerequisite for static
stripe order which may compete with superconductivity in
cuprates.\cite{tranquada95,grafe06,vojta09}

In view of the intimate coupling of the respective structure to
superconductivity, a detailed understanding of the \textit{local}
structure and its doping- and temperature-dependence is crucial to
understand the high-$T_c$ mechanism. Yet, there have been
unsettled issues regarding the local structure such as whether the
macroscopic average structure fully corresponds to the local one
involving a displacive structural
transition,\cite{friedrich96,braden01,simovic03a} or to a
superposition of coherent local LTT variants involving an
order-disorder
transition.\cite{billinge94,haskel96,haskel00,han02}

While nuclear magnetic resonance (NMR) as a local probe is
suitable to investigate the local structure, it is also extremely
sensitive to inhomogeneities in the sample which often complicate
the NMR spectra, leading to an erroneous analysis of the results.
For example, the magnetic phase separation
scenario\cite{julien99a} in LSCO with $x=0.06$ based on the
splitting of the $^{139}$La NMR spectrum turned out to be a
misinterpretation,\cite{julien01} although its true origin has not
been thoroughly understood.

In this paper, the local structure in a high quality \lscos\
single crystal is investigated via $^{139}$La NMR. The La site is
most strongly affected by the local distortions associated with
the structural phase transition,\cite{radaelli94,haskel96} making
the $^{139}$La a suitable local probe for the local structure.
Indeed, our $^{139}$La NMR results show that the local tilting
pattern of the CuO$_6$ octahedra in the LTO structure is
accompanied by a tilting of the EFG at the nuclei, resulting in
the splitting of the otherwise single resonance line. We confirmed
that the $^{63}$Cu spectrum splits into two in the LTO structure
as well, but with a much smaller tilt angle than the value for the
$^{139}$La. A strong temperature dependence was found in the
splitting of the $^{139}$La spectrum, which measures the tilt
angle of the CuO$_6$ octahedra which disappears abruptly above
$T_S$. Our NMR data thus provide an evidence that the macroscopic
average structure parameters are sufficient to account for the
local lattice distortions and their temperature dependence. In
addition, the $^{139}$La spin-lattice relaxation rates measured in
a wide temperature range are discussed.

\section{Sample preparation and Experimental details}

The crystal of \lscos\ ($T_c=14$ K) has been grown from 4N materials of
\ce{La2O3}, \ce{SrCO3}, and CuO using the TSFZ (Traveling Solvent Floating Zone)
technique. Using this technique, which is described in detail elsewhere,\cite{lambacher10}
large crystals of several centimeters length can be grown
under accurately controllable stable conditions (flux composition, temperature
and oxygen partial pressure), which in terms are the prerequisite for
homogeneous crystals.
The sample for the NMR measurements was cut out of an single crystalline rod
after the growth procedure along the crystallographic $a$, $b$, and $c$-axis using
the  Laue x-ray  method. The outer dimensions measured 1.7 mm,
2.5 mm, and 0.8 mm in $a$, $b$, and $c$-direction.

$^{139}$La ($I=7/2$) and $^{63}$Cu ($I=3/2$) NMR experiments were
performed on the single crystal of \lscos\ in the range of
temperature 4.2--420 K.  The sample was mounted on
a goniometer which allows an accurate alignment of the sample along the
external field. In
order to check the quadrupole frequency $\nu_Q$, the $^{139}$La
NQR spectrum at the $3\nu_Q$ ($\pm\frac{7}{2}\leftrightarrow \pm\frac{5}{2}$)
transition ($\sim18.9$ MHz) and the $^{63}$Cu NQR
spectrum at the $\nu_Q$ ($\pm\frac{3}{2}\leftrightarrow \pm\frac{1}{2}$)
transition ($\sim34$ MHz) were measured. The NMR/NQR spectra were
acquired by spin-echo signals as the frequency was swept through the resonance line using
Hahn echo sequence ($\pi/2-\tau-\pi$) with $\pi/2\sim2$ $\mu$s.

The nuclear spin-lattice relaxation rates \slr\ were measured at the central
transition of the $^{139}$La by the saturation recovery method and the relaxation
of the nuclear magnetization after a saturating pulse was fitted to
the following equation:
\begin{equation}
\begin{split}
\label{eq:T1}
1-\frac{M(t)}{M(\infty)}=
a&\left(\frac{1}{84}e^{-t/T_1}+\frac{3}{44}e^{-6t/T_1}\right.   \\
&+ \left.\frac{75}{364}e^{-15t/T_1}+\frac{1225}{1716}e^{-28t/T_1}\right),
\end{split}
\end{equation}
where $M$ is the nuclear magnetization and $a$ a fitting parameter that is
ideally one.

\section{Results and Discussion}

\subsection{$^{139}$La NMR spectra}

\begin{figure}
\centering
\includegraphics[width=\linewidth]{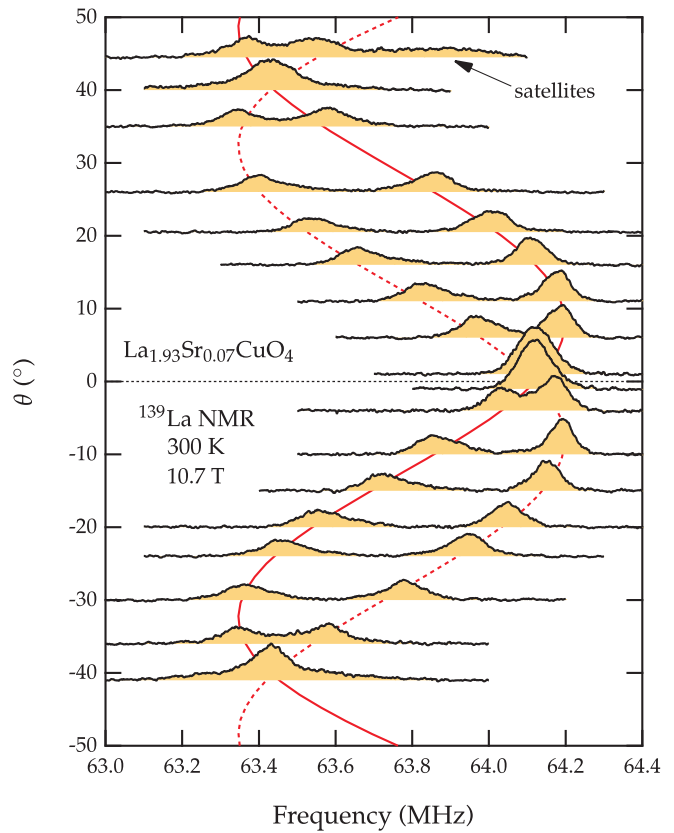}
\caption{\label{fig:lasp1} Angle dependence of $^{139}$La NMR spectra
measured at room temperature and 10.7 T.  Solid
(dotted) lines  are theoretical calculations assuming $8^\circ$ ($-8^\circ$)
tilting of the principal axis of the EFG from the $c$ axis, using
$\mathcal{K}=-0.09$ \% and $\nu_Q=6.3$ MHz which were determined experimentally from this
study.  The anisotropy parameter $\eta=0.19(2)$ was obtained from the calculations.}
\end{figure}

Figure \ref{fig:lasp1} shows the central transition NMR spectrum of
the $^{139}$La at 300 K and 10.7 T as a function of the angle
$\theta$ between the crystallographic $c$ axis and the external
field $H$. The $^{139}$La spectrum consists of two resonance lines
which both exhibit the same angle dependence with respect to $H$:
the peaks are centered at $\theta_m^\pm=\pm8^\circ$, respectively,
where the maximum resonance frequencies are observed. The two
angle-dependent sets of the $^{139}$La spectrum which are
symmetric around the $c$ axis imply the existence of two different
sites with the same occupation probability.

Such a strong angle dependence of the central
transition is expected from
second order quadrupole interaction of the nuclear quadrupole moment $Q$ that
is non-zero for the nuclear spin $I>\frac{1}{2}$, with
the EFG. For the
central transition ($-\frac{1}{2}\leftrightarrow \frac{1}{2}$), while the 
first order quadrupole shift is zero, the second
order quadrupole shift is given by\cite{bennet}
\begin{equation}
\begin{split}
\label{eq:quad}
      \Delta\nu^\text{2nd} = \frac{\nu_Q^2}{32\gamma_n H}
      (1-&\cos^2\theta)\left[ 2I(I+1)-\frac{3}{2}\right.   \\
      &-\left. \left(18I(I+1)-\frac{27}{2}\right)\cos^2\theta \right],
\end{split}
\end{equation}
where $\gamma_n$ is the nuclear gyromagnetic ratio and $\nu_Q$ is the
quadrupole frequency given by $3e^2qQ/2I(2I-1)h$ where $eq$ is the EFG. Here,
without the loss of generality,
we ignored an anisotropy parameter $\eta$ which is non-zero in sites of
lower than axial symmetry.
From Eq.~(\ref{eq:quad}), one can see that when $\theta=0$, $\Delta\nu=0$ so
that the resonance frequency becomes one of the maxima and directly yields the Knight
shift since a quadrupole correction is not needed.

\begin{figure}
\centering
\includegraphics[width=\linewidth]{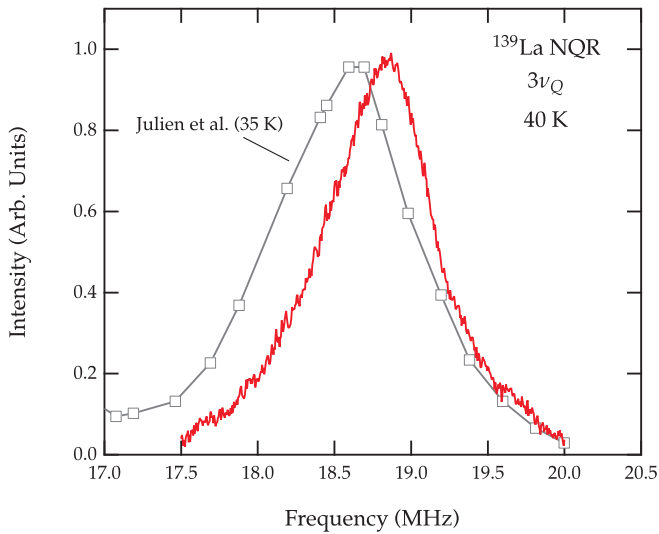}
\caption{\label{fig:lanqr} $^{139}$La NQR spectrum measured at the $3\nu_Q$
transition at 40 K giving rise to $\nu_Q=6.3$ MHz.
The NQR spectrum from Julien et al.\cite{julien01}
measured in the sample of $x=0.1$ is compared. The narrower linewidth of our
NQR spectrum indicates less chemical inhomogeneity in the single crystal.  }
\end{figure}

It may be possible that the $^{139}$La is differentiated due to the presence of two EFG
strengths which would cause the different quadrupole shift at the nuclei. To check
the possibility, we measured the $^{139}$La NQR spectrum in zero field at the
$3\nu_Q$ transition.
Since a well-defined single NQR line as shown in Fig.~\ref{fig:lanqr} indicates
the unique value of the EFG in the system,
the two $^{139}$La peaks are attributed to the local tilting of
the direction of the EFG at the $^{139}$La
with respect to the $c$ axis,\cite{hammel93}  where $\theta_m$ corresponds to the
angle at which the direction of the EFG is parallel to
$\mathbf{H}$.

In order to confirm this quantitatively, we performed exact
diagonalization calculations of the nuclear Hamiltonian which can
be written as,

\begin{equation}
\label{eq:Ham}
\begin{split}
    \mathcal{H} =& -\gamma_n\hbar[1+\mathcal{K}(\theta)]\mathbf{I}\cdot\mathbf{H} \\
    & + \frac{\nu_Q}{6} \left[3I_z^2-I^2+\eta \left(I_x^2-I_y^2\right)\right],
\end{split}
\end{equation}
where $\mathcal{K}$ is the Knight
shift which is usually angle dependent and $I$ is the nuclear spin operator.

In the calculations, we used $\nu_Q=6.3$ MHz from the
$3\nu_Q$ transition (Fig.~\ref{fig:lanqr}) and $\mathcal{K}=-0.09$ \%
(Fig.~\ref{fig:freq}). Here we neglect a possible angle dependence of
$\mathcal{K}$. Thus, $\eta$ is the only unknown
parameter in the Hamiltonian. By adjusting $\eta$, we accurately reproduced the resonance
frequencies of the spectra as a function of $\theta$, drawn as solid (dotted) lines in
Fig.~\ref{fig:lasp1}. From these calculations, the anisotropy parameter $\eta=0.19(2)$
was deduced. Since $\eta$ can absorb the effect
of the angle variation of $\mathcal{K}$, the actual value may slightly differ
from the one deduced here.
The excellent agreement of the theory with the experimental data
indicates a staggered pattern of the tilting of the EFG at the $^{139}$La
around the $c$ axis, due to the corresponding local structural distortions.

\begin{figure}
\centering
\includegraphics[width=\linewidth]{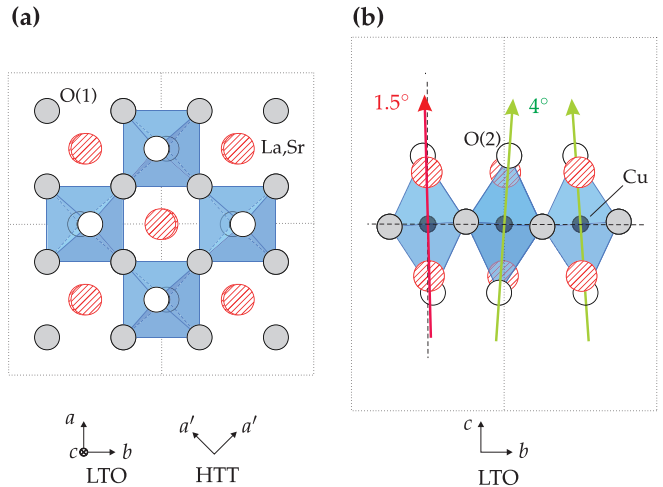}
\caption{\label{fig:structure} Structure of \lsco\ in the LTO phase at 10 K
for $x=0.075$, using the structure parameters reported in
Ref.~\onlinecite{radaelli94}, viewed
along [001] (a) and
along [100] (b). The unit cell is depicted as the dotted line.
The CuO$_6$ octahedra is tilted along
the [010]$_\text{LTO}$ (or [110]$_\text{HTT}$) direction by $4^\circ$.
In addition, La(Sr)-La(Sr) along the $c$ axis is tilted by $1.5^\circ$ in
opposite direction of the distortion of the \ce{CuO6} octahedra, being consistent
with a much larger tilted angle of the EFG at the $^{139}$La.
Note that the tilt direction of both the \ce{CuO6} octahedra and the La-La is maintained
along [010]$_\text{LTO}$,
whereas, along the diagonal direction i.e., [110]$_\text{LTO}$, it is rotated
alternately by 180$^\circ$, causing the so called \textit{buckling} of
the CuO$_2$ plane.  }
\end{figure}

As a matter of fact, this is well supported by the LTO structure
in Fig.~\ref{fig:structure}, which was drawn using structure
parameters for $x=0.075$ at 10 K given in
Ref.~\onlinecite{radaelli94}. Clearly, the main axis of the
CuO$_6$ octahedra is tilted by $4^\circ$ along the [010] direction
(or [110] in the HTT setting). In addition, the position of La(Sr)
with respect to the $c$ axis is slightly shifted in opposite
direction of the distortion of the CuO$_6$ octahedra, which may
cause a larger tilting of the EFG at the $^{139}$La than at the
$^{63}$Cu. Furthermore, the fact that the tilt direction of the
octahedra is reversed alternately along the [110]$_\text{LTO}$,
which induces the buckling of the CuO$_2$
planes,\cite{axe94,tranquada95} accounts for the two $^{139}$La
peaks whose intensities and positions as a function of $\theta$
are symmetric around the $c$ axis.

\begin{figure}
\centering
\includegraphics[width=\linewidth]{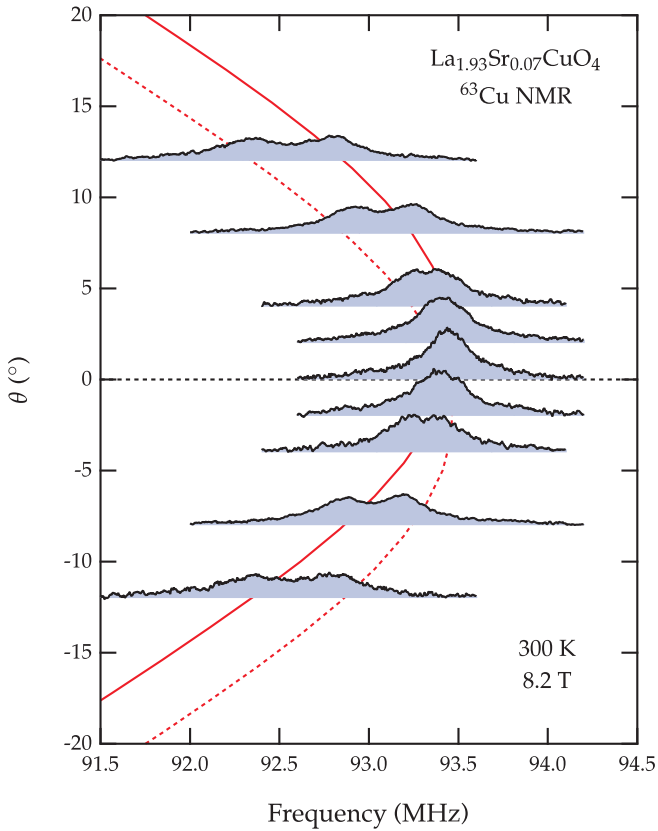}
\caption{\label{fig:cusp} Angle dependence of the $^{63}$Cu NMR spectra
measured at 300 K and 8.2 T.  The
solid (dotted) lines are theoretical calculations assuming $2^\circ$
($-2^\circ$) tilting of the direction of the EFG at the $^{63}$Cu from the $c$
axis, with parameters $\nu_Q=34$ MHz, $K=1.18$ \%, and $\eta=0$.}
\end{figure}

If the two $^{139}$La NMR lines are the result of the distorted
CuO$_6$ octahedra in the LTO phase, one should expect that there
are two Cu sites which are distinguishable in field as well.
However, the presence of the two central transitions of the
$^{63}$Cu has not been reported so far. It may be possible that
the tilt angle of the EFG at the $^{63}$Cu is too small to be
detected, especially with the relatively larger linewidth of the
$^{63}$Cu NMR spectrum than that of the $^{139}$La. In order to
check whether there exist two $^{63}$Cu NMR lines in the LTO
phase, the $^{63}$Cu spectrum has been carefully examined by
varying $\theta$ at 300 K. Indeed, as shown in
Fig.~\ref{fig:cusp}, the angle dependence of the $^{63}$Cu
spectrum unravels two central transitions which are centered at
$\theta_m^\pm = \pm 2^\circ$ respectively, being symmetric around
the $c$ axis. The solid (dotted) lines are from theoretical
calculations with Eq.~(\ref{eq:Ham}) using the values of
parameters, $\nu_Q=34$ MHz and $\mathcal{K}_c=1.18$ \%. Unlike the
case of the $^{139}$La, a non-zero $\eta$ is not needed to fit the
data. The negligible $\eta$ and the much smaller $\theta_m$ than
that of the $^{139}$La are also consistent with the LTO structure
depicted in Fig.~\ref{fig:structure}, where the La itself is
displaced. Therefore, we conclude that the two peaks at both the
$^{139}$La and the $^{63}$Cu are a natural consequence of the
HTT-LTO structural transition in \lsco.

We are aware that the qualitatively similar argument was given by
Julien \textit{et al}.\cite{julien01} where they observed four
$^{139}$La peaks which were attributed to the tilting of the
direction of the EFG about the $c$ axis.  However, the four peaks
with various signal intensities and linewidths are difficult to
understand in detail and are not in accordance with only two
different La sites as expected for the LTO structure. We
conjecture that the sample used by Julien \textit{et al}.
consisted of two or more domains which have slightly different
orientations.

\subsection{$^{139}$La Knight shift and structural order parameter}

Next we discuss the temperature dependence of the $^{139}$La
spectra. In order to extract the Knight shift for $I>1/2$ in
non-axial symmetry, the second order quadrupole effect which
shifts the central transition, as given by Eq.~(\ref{eq:quad}),
should be taken into careful consideration, particularly for a
very small Knight shift which is the case for the $^{139}$La NMR
in \lsco. Taking advantage of the fact that the second order
quadrupole shift vanishes when $H$ is applied along the direction
of the EFG, we deliberately rotated the sample with respect to the
$c$ axis by $8^\circ$. Then, the second order quadrupole shift for
the peak p1 is zero and the intrinsic Knight shift as a function
of temperature could be extracted.

\begin{figure}
\centering
\includegraphics[width=\linewidth]{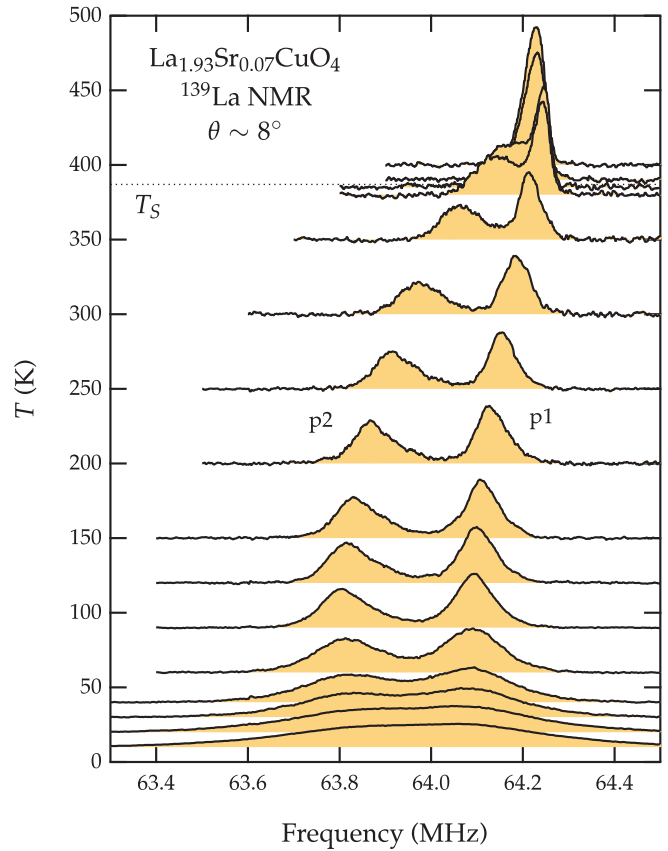}
\caption{\label{fig:lasp2} Temperature-dependence of the $^{139}$La NMR
spectra in an external field tilted by $\theta\sim8^\circ$ from the
crystallographic $c$
axis. In this case, p1 corresponds to the line in which the EFG at the
$^{139}$La is aligned along $H$, giving rise to the Knight shift for
$H\parallel c$ since there is no quadrupole shift. At 387 K, the two peaks
collapse to a single peak, indicating the LTO-HTT structural transition. At
low temperatures,
$^{139}$La spectrum is significantly broadened, implying the slowing down of the
spin fluctuations due to the spin-density wave (SDW) order. Each spectrum was
scaled with temperature to remove the Boltzmann contribution to the intensity.}
\end{figure}

\begin{figure}
\centering
\includegraphics[width=\linewidth]{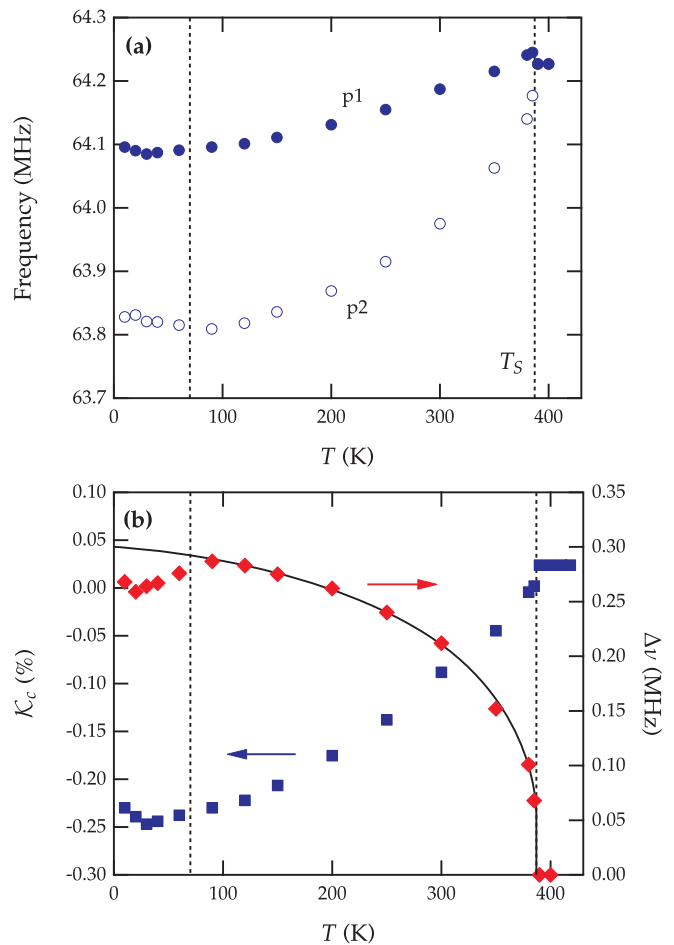}
\caption{\label{fig:freq} (a) Temperature-dependence of resonance frequencies of
two NMR peaks in Fig. \ref{fig:lasp2}. (b) Knight shift of the $^{139}$La is extracted
from the p1 line. The separation of the two peaks $\Delta\nu(T)$
behaves as a structural order parameter. $\Delta\nu$ reveals first order-like
discontinuous transition at
$T_S=387$ K, while it exhibits the second order-like temperature dependence except the region
of the transition. Solid line is a guide to eyes. Below 70 K, data deviate due to the
onset of the magnetic broadening associated with the incommensurate SDW order.}
\end{figure}

With increasing temperature from 300 K, the separation between the
two peaks becomes smaller and the spectra collapse abruptly into
single spectrum at $T_S$ = 387(1) K, the structural transition
temperature $T_S$ from the HTT to the LTO
phase.\cite{keimer92,nakano94} This is another direct evidence
that the two $^{139}$La central peaks arise from the tilted
CuO$_6$ octahedra associated with the LTO structure. The
$T$-dependence of the resonance frequencies of the two lines is
shown in Fig.~\ref{fig:freq}(a). The single line above $T_S$ is
split into two at $T_S$, in which one line (p1) shifts up and the
other (p2) shifts down in frequency from the single line above
$T_S$. The line p1 gives rise to the Knight shift for $H\parallel
c$, $\mathcal{K}_c$, which is drawn in Fig.~\ref{fig:freq}(b).
Note that the Knight shift data above $T_S$ is obtained after
rotating back the sample along the $c$ axis, since the EFG should be
directed along the crystallographic $c$ axis in the HTT structure due to 
the axial symmetry with respect to the $c$ axis.

Additional valuable information was obtained from
Fig.~\ref{fig:freq}(a). Just below $T_S$, the separation between
p1 and p2 rapidly increases but approaches a constant value at
low temperatures, that is, the separation between the two lines
$\Delta\nu$ as a function of temperature behaves as the structural
order parameter as shown in Fig.~\ref{fig:freq}(b). At $T_S$, the
transition is clearly discontinuous, indicating a first order
nature of the transition, in contrast to second-order character
reported in the diffraction measurement.\cite{axe94} On the other
hand, except for the discontinuous change at the transition, the
temperature dependence of $\Delta\nu$ shows the behavior expected
in a second order transition.

The narrow single resonance line in the HTT phase indicates that
within the resolution of the NMR experiment the local tilting of
the CuO$_6$ octahedra vanishes immediately above $T_S$ rendering
the local tetragonal symmetry, in contrast to the experimental
reports that the LTO-type local tilt remains even in the HTT
phase.\cite{haskel96,wakimoto06} Furthermore, the abrupt splitting
of the spectrum at $T_S$ confirms a displacive transition, not an
order-disorder one.


Since $\Delta\nu$ is a measure of the tilt angle of the EFG at the
nuclei, its rapid suppression with increasing temperature from 300
K toward $T_S$ may cause a finite second order quadrupole shift
affecting the $^{139}$La Knight shift which was obtained with the
tilt angle $8^\circ$ determined at 300 K. Then, the actual Knight
shift data just below $T_S$ may be slightly enhanced, in such a
way that data are smoothly connected to those in the HTT phase.
Likewise, $\Delta\nu$ itself should be affected with a small
error.

\subsection{$^{139}$La spin-lattice relaxation rate}

\begin{figure}
\centering
\includegraphics[width=\linewidth]{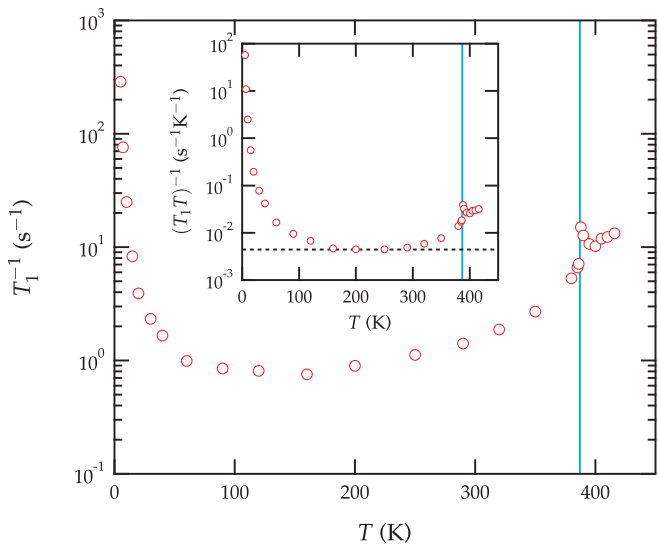}
\caption{\label{fig:T1} $^{139}$La \slr\ as a function of temperature.  A
very sharp anomaly at $T_S=387$ K was detected with the transition width less
than 10 K. Below 70 K, the critical slowing down due to the SDW order causes
extremely steep upturn of \slr.  Inset shows \slrt\ of the $^{139}$La versus temperature.
Note that \slrt\ increases more than four decades as the temperature increases
from $\sim100$ K down to 4.2 K.}
\end{figure}

Figure \ref{fig:T1} shows the spin lattice relaxation rate \slr\ of
the $^{139}$La as a function of temperature in the range 4.2--420
K. With increasing temperature from room temperature, \slr\ is
rapidly enhanced. We found a discontinuous jump at $T_S=387$ K
which is followed immediately by a steep drop above $T_S$. The
sharp anomaly of \slr\ around $T_S$ is consistent with the drastic
change of the $^{139}$La spectrum at $T_S$ (Fig.~\ref{fig:lasp2})
and supports a first order structural transition. Upon further
increasing the temperature, \slr\ continues to increase up to 420
K which is a limitation of our equipment.

Interestingly, the temperature dependence of $^{139}T_1^{-1}$ resembles that
observed in \lesco,\cite{suh99,curro00} where
$^{139}T_1^{-1}$ rapidly drops below
the LTT-LTO structural transition. The
HTT-LTO transition, however, is clearly manifested through an anomaly in the narrow
range of temperature
which is simply added on top of the rapid drop of \slr. Note that \slr\
already started to decrease at a much higher temperature than $T_S$.

Rather, we conjecture that the rapid upturn of \slr\ is associated
with a gradual change from a system with a local moment character to a
more itinerant system, as argued by Imai \textit{et al}.\cite{imai93} They
report that $^{63}T_1^{-1}$ of the $^{63}$Cu forms a broad maximum
near 200 K in similarly Sr-doped samples, approaching a
temperature independent value $^{63}T_{1\infty}^{-1}\sim2500$
s$^{-1}$ at sufficiently high
temperatures.\cite{imai93,fujiyama97} If the temperature and
doping independent $^{63}T_{1\infty}^{-1}$ at $T\sim
J_\text{ex}/2$, where $J_\text{ex}$ is the exchange coupling constant,
imply the exchange narrowed paramagnetic limit, one
may expect that $^{139}T_{1\infty}^{-1}$ should be dominated by
$J_\text{ex}$ and the hyperfine coupling
$A_\text{hf}$ according to the
relation,\cite{moriya56a,chakravarty90}
\begin{equation}
\label{moriya}
 \frac{1}{T_{1\infty}} \propto \frac{A_\text{hf}^2 \sqrt{S(S+1)}}{J_\text{ex}},
\end{equation}
where we neglect an anisotropy of $A_\text{hf}$ for simplicity. A
very rough estimation from the Knight shift values, which are
proportional to $A_\text{hf}$, obtained at room temperature gives
$^{139}T_{1\infty}^{-1} = {^{63}T}_{1\infty}^{-1}
(^{139}\mathcal{K}/^{63}\mathcal{K})^2\sim 10$ s$^{-1}$ which is
surprisingly close to the experimental values near 400 K. Then,
one may anticipate that with increasing temperature beyond 420 K
$^{139}T_1^{-1}$ increases further and decreases ultimately, or
gradually saturates, to a constant value that is close to $\sim
10$ s$^{-1}$, being temperature-independent. 
This indicates that the $^{139}$La relaxation rates at high temperatures 
near $J_\text{ex}/2$ are governed by the same local spin dynamics probed in the $^{63}$Cu 
results.\cite{imai93} Therefore, the $S=\frac{1}{2}$
antiferromagnetic 2D Heisenberg model\cite{chakravarty90,gelfand93} may be indeed 
appropriate to describe the high temperature spin dynamics in LSCO even at 
$T\sim J_\text{ex}/2$.  

Furthermore, we find that \slr\ decreases linearly with
temperature below 300 K down to $\sim150$ K. This is clearly shown
by \slr\ divided by temperature, \slrt\ (see the inset of
Fig.~\ref{fig:T1}). The Korringa-like behavior suggests that the
relaxation process of the $^{139}$La is dominated by the doped
hole carriers in this temperature range. Therefore, we ascribe the
rapid drop of \slr\ at high temperatures to an evolution from a
paramagnetic regime at $T\ge J_\text{ex}/2$ to a regime below 300
K where the scattering with doped holes dominates the spin
fluctuation properties, and thus the $^{139}$La \slr.

Below $\sim70$ K, \slr\ is enhanced in an extremely fast rate,
which is ascribed to the critical slowing down of spin
fluctuations associated with glassy spin freezing.  Such a rapid
upturn of \slr\ is compatible with the significant broadening of
the $^{139}$La NMR spectrum in the same temperature range as shown
in Fig.~\ref{fig:lasp2}. Down to 4.2 K, a maximum of \slr, which
gives a characteristic spin freezing temperature $T_g$, was not
identified. This is somewhat inconsistent with previous $^{139}$La
NMR results,\cite{julien99a,julien01} and suggests a reduced $T_g$
which is possibly related to less disorder in our sample.

\section{Conclusion}

Through $^{139}$La and $^{63}$Cu NMR, we proved that the local
tilting of the CuO$_6$ octahedra along [110]$_\text{HTT}$ occurs
in the LTO phase, whereas it is absent in the HTT phase in \lscos.
The tilt direction of the CuO$_6$ octahedra alternates along the
same direction resulting in the buckling of the CuO$_2$ plane. The
structural order parameter representing the tilt angle of the
CuO$_6$ octahedra was successfully obtained from the splitting of
the $^{139}$La resonance line as a function of temperature. The
structural transition at $T_S=387$ K is weakly first order and
displacive i.e., the order parameter is reduced rapidly with
increasing temperature approaching $T_S$, and abruptly vanishes in
the HTT phase. In other words, the average HTT as well as the LTO
phases are fully compatible with their local structures. Our data
also rule out the possibility of an inhomogeneous mixture of the
two phases in the whole temperature range measured in this study.

The $^{139}$La spin-lattice relaxation rates, \slr, exhibits a
sharp anomaly at $T_S$, which is consistent with the abrupt change
of the spectrum. The temperature dependence of \slr\ in the high
temperature region suggests that a
temperature-independent paramagnetic regime at high temperatures changes 
gradually to a regime where the
scattering with doped holes dominates the $^{139}$La \slr. The
strong enhancement of \slr\ below $\sim70$ K is ascribed to the
critical slowing down of spin fluctuations associated with glassy
spin freezing.


\section*{Acknowledgement}

The authors wish to thank Markus H\"ucker for his helpful
discussions and communications. This work has been supported by
the DFG through FOR 538 (Grant No. BU887/4 and No. ER342/1-3) and
SPP1458 (Grant No. GR3330/2).

\bibliography{mybib}

\end{document}